\documentclass{article}

\usepackage{PRIMEarxiv}

\usepackage[utf8]{inputenc} 
\usepackage[T1]{fontenc}    
\usepackage{hyperref}       
\usepackage{url}            
\usepackage{booktabs}       
\usepackage{amsfonts}       
\usepackage{nicefrac}       
\usepackage{microtype}      
\usepackage{lipsum}
\usepackage{fancyhdr}       
\usepackage{graphicx}       
\graphicspath{{media/}}     

\usepackage{xspace}
\usepackage{enumitem}
\usepackage{multirow}
\usepackage{balance}
\usepackage{booktabs}
\usepackage{url}

\newcommand{\method}{DriveRLR\xspace}

\newcommand{\gpt}{\textit{GPT-5}\xspace}
\newcommand{\llama}{\textit{Llama 4 Maverick}\xspace}
\newcommand{\mistral}{\textit{Mistral Small 3.2}\xspace}

\title{A Tool for Benchmarking Large Language Models' Robustness in Assessing the Realism of Driving Scenarios}

\author{
  Jiahui Wu \\
  Simula Research Laboratory and \\ University of Oslo \\
  Oslo, Norway \\
  \texttt{jiahui@simula.no} \\
   \And
  Chengjie Lu \\
  Simula Research Laboratory and \\ University of Oslo \\
  Oslo, Norway \\
  \texttt{chengjielu@simula.no} \\
   \And
  Aitor Arrieta \\
  Mondragon University \\
  Mondragon, Spain \\
  \texttt{aarrieta@mondragon.edu} \\
   \And
  Shaukat Ali \\
  Simula Research Laboratory \\
  Oslo, Norway \\
  \texttt{shaukat@simula.no} \\
}

\begin{document}
\maketitle

\begin{abstract}
In recent years, autonomous driving systems have made significant progress, yet ensuring their safety remains a key challenge. To this end, scenario-based testing offers a practical solution, and simulation-based methods have gained traction due to the high cost and risk of real-world testing. However, evaluating the realism of simulated scenarios remains difficult, creating demand for effective assessment methods.
Recent advances show that Large Language Models (LLMs) possess strong reasoning and generalization capabilities, suggesting their potential in assessing scenario realism through scenario-related textual prompts. 
Motivated by this, we propose \method, a benchmark tool to assess the robustness of LLMs in evaluating the realism of driving scenarios. \method generates mutated scenario variants, constructs prompts, which are then used to assess a given LLM's ability and robustness in determining the realism of driving scenarios.
We validate \method on the DeepScenario dataset using three state-of-the-art LLMs: \gpt, \llama, and \mistral. Results show that \method effectively reveals differences in the robustness of various LLMs, demonstrating its effectiveness and practical value in scenario realism assessment.
Beyond LLM robustness evaluation, \method can serve as a practical component in applications such as an objective function to guide scenario generation, supporting simulation-based ADS testing workflows.
\end{abstract}

\keywords{Realistic driving scenarios \and Large language models \and Robustness}

\section{Introduction}
\noindent
Autonomous Driving Systems (ADSs) have been advancing rapidly, with extensive research and development~\cite{zhao2024autonomous}. However, ensuring their safety remains the key to strengthening public confidence in their adoption. To guarantee ADS safety in dynamic and complex environments, testing is particularly crucial~\cite{tang2023survey}. 
Scenario-based testing, which can generate both typical and extreme driving scenarios, has emerged as an effective approach for ADS testing~\cite{zhang2022finding,nalic2020scenario}. 
However, real-world scenario testing is often costly, and certain high-risk scenarios are difficult or even impossible to safely produce in real-world. As a result, simulation-based testing provides a feasible and effective alternative~\cite{nalic2020scenario}. 

In recent years, various simulation-based testing methods have been proposed, demonstrating clear advantages~\cite{calo2020generating,lu2024epitester,feng2023dense}. However, discrepancies exist between real-world and simulated scenarios~\cite{stocco2022mind}. For example, simulated environments struggle to accurately reproduce the dynamic trajectories of vehicles. 
Although some studies have attempted to mitigate this gap via optimizing ADS configurations to avoid problematic scenarios~\cite{calo2020generating}, enforcing realistic constraints during scenario generation~\cite{lu2023deepqtest}, and leveraging real-world driving data~\cite{yan2023learning}, these solutions remain limited. Thus, the realism of test scenarios is still difficult to guarantee~\cite{liao2025advancing}.
Therefore, effectively evaluating the realism of simulated scenarios has become a critical challenge that urgently needs to be addressed.

Large Language Models (LLMs) have demonstrated near or even human-expert-level performance in tasks such as commonsense reasoning and multiple-choice question answering~\cite{achiam2023gpt}. They can effectively understand textual information, uncover underlying logical patterns, and make reasonable inferences.
Motivated by this, Wu et al.~\cite{wu2024reality} hypothesize that LLMs have the potential to assess the realism of driving scenarios. To validate this hypothesis, they conducted an empirical study to evaluate the robustness of LLMs in this context, and designed a novel metric to quantify their robustness in scenario realism assessment. Building on this evidence, we introduce \method, a benchmark tool for evaluating the robustness of LLMs in the realism assessment of driving scenarios. 
\method applies small mutations to original driving scenarios to create variants, generates corresponding natural language prompts, and submits them to LLMs for realism evaluation. By analyzing the variation in outputs across these scenario variants, \method enables systematic assessment of LLM robustness in scenario realism.

\method offers three implementation modes tailored to different user needs, i.e., a terminal-based tool, a Python package for function-level control, and full source code for advanced customization and extension.
Furthermore, we use driving scenarios from DeepScenario~\cite{lu2023deepscenario} as realistic scenarios to test the implementation of the tool, as they incorporate realistic constraints in scenario generation. The assessment was conducted using the latest three LLMs, \gpt, \llama, and \mistral, and the results demonstrate the effectiveness and practical value of \method. Furthermore, \method is valuable for benchmarking existing or new LLMs to assess their ability to determine realism, which can then be used by simulation-based ADS testing techniques, for example, by serving as an objective function to guide scenario generation or identifying high-realism scenarios within a dataset. Ultimately, \method contributes to more efficient, targeted, and reliable safety validation of ADSs.

\section{\method Tool}
\begin{figure}[!t]
    \centering
    \includegraphics[width=0.8\linewidth]{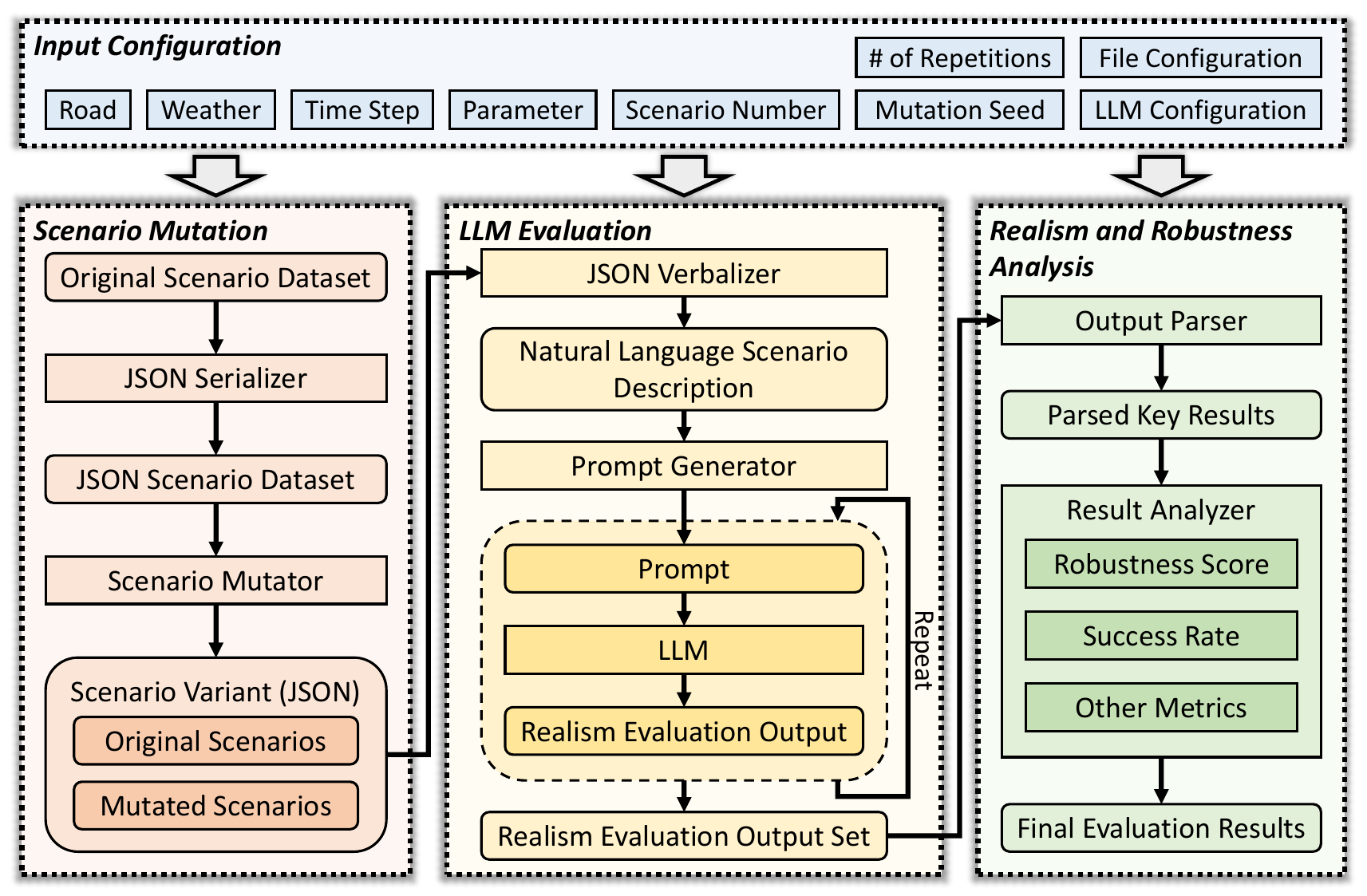}
    \caption{Overview of \method.}
    \label{fig:overview}
\end{figure}

\noindent
Figure~\ref{fig:overview} illustrates the overview architecture of \method, which consists of four main modules: \textit{Input Configuration}, \textit{Scenario Mutation}, \textit{LLM Evaluation}, and \textit{Realism and Robustness Analysis}.
\textit{Input Configuration} module provides the necessary parameter settings for benchmark execution and ensures standardized input formats and control settings for the remaining modules.
\textit{Scenario Mutation} module preprocesses the original driving scenarios by applying small numerical perturbations to scenario parameters, thereby generating a set of variant scenarios, which are used to assess the robustness of LLMs in evaluating scenario realism.
\textit{LLM Evaluation} module generates corresponding prompts based on the original and mutated scenarios, guiding LLMs to perform realism analysis and produce evaluation outputs for each scenario variant.
\textit{Realism and Robustness Analysis} module integrates a set of evaluation metrics, including the robustness indicator and realism success rate, to comprehensively analyze the LLMs' performance in scenario realism evaluation and their robustness across different scenario variants.
In this section, we provide a detailed description of the functionality and workflow of each module in \method.

\subsection{Input Configuration}
\noindent
\textit{Input Configuration} module includes various configurable parameters, such as road types, weather conditions, and LLM settings, as shown in Figure~\ref{fig:overview}. It provides a standardized parameter format and input control mechanism, which facilitates systematic workflow management within \method and ensures the reproducibility and consistency of subsequent experiments. A detailed description of the parameters in this module will be provided in Section~\ref{subsec:Configuration}.

\subsection{Scenario Mutation}
\noindent
\textit{Scenario Mutation} module is responsible for applying minor perturbations to the original scenarios under evaluation. While ensuring that the mutated scenarios remain realistic, it generates a set of scenario variants for subsequent analysis of LLM robustness in scenario realism assessment.
As shown in Figure~\ref{fig:overview}, \method first loads the \textit{Original Scenario Dataset} into the \textit{JSON Serializer}, converting it into a standardized \textit{JSON Scenario Dataset}, which is then passed to the \textit{Scenario Mutator} module.
Based on the specific parameters specified in the \textit{Input Configuration}, \textit{Scenario Mutator} module applies small numerical mutations to the corresponding elements within each scenario.
The extent of each mutation is controlled by a configurable value called the \textit{Mutation Seed}, which is also specified in the \textit{Input Configuration}, allowing for flexible adjustment.
Taking the mutation of non-player character (NPC) vehicle positions as an example, \method allows users to specify the corresponding parameter name in the \textit{Input Configuration} (e.g., position, which must match the field name in the JSON scenario), enabling the \textit{Scenario Mutator} to automatically locate and modify the position value based on the specified \textit{Mutation Seed}.
For instance, when the \textit{Mutation Seed} is set to 0.01, the position value increases by 1\%, i.e., mutated position = original position × 1.01, resulting in a subtly perturbed scenario.
By flexibly configuring different scenario parameters, \method can generate multiple mutated versions of each original scenario. These mutated scenarios, together with the originals, form a complete set of scenario variants, which are then used in the \textit{LLM Evaluation} module.

\subsection{LLM Evaluation}
\noindent
The main task of the \textit{LLM Evaluation} module is to generate prompts for each scenario variant and use them to guide LLMs in assessing scenario realism, producing corresponding evaluation outputs.
In this module, \method first feeds the preprocessed scenario variants into the \textit{JSON Verbalizer}, which converts them into \textit{Natural Language Scenario Description}. These descriptions, combined with parameters from the \textit{Input Configuration}, such as road type, weather conditions, and time steps, are passed to the \textit{Prompt Generator} to create prompt instances suitable for LLM input.
To ensure accuracy and consistency, each prompt is submitted to the LLM multiple times, generating several \textit{Realism Evaluation Output}. These outputs collectively form a \textit{Realism Evaluation Output Set}, which is then used for subsequent \textit{Realism and Robustness Analysis}.

\subsection{Realism and Robustness Analysis}
\noindent
\textit{Realism and Robustness Analysis} module analyzes the evaluation results produced by LLMs and performs quantitative analysis based on metrics such as \textit{Robustness Score} and \textit{Realism Success Rate}, as proposed in~\cite{wu2024reality}, to generate the final evaluation results.
In this module, \method first receives the \textit{Realism Evaluation Output Set} from the \textit{LLM Evaluation} and processes it through the \textit{Output Parser} to extract \textit{Parsed Key Results}, e.g., whether the scenario is realistic. These intermediate results are then quantitatively analyzed using the metrics defined in~\cite{wu2024reality}, resulting in the \textit{Final Evaluation Results}.
Moreover, parameters such as road type and weather conditions, provided in the \textit{Input Configuration}, enable more detailed and specific evaluation and result interpretation.

\section{Tool Usage}
\noindent
To use \method, users can configure parameters based on their specific needs. \method offers three usage modes to accommodate different types of users with varying functional requirements.
This section provides a detailed explanation of both aspects, along with an introduction to the intermediate outputs and final evaluation results generated by \method.

\subsection{Configuration}\label{subsec:Configuration}
\noindent
\method offers a set of configurable parameters to coordinate the execution of subsequent modules, as shown in the \textit{Input Configuration} module in Figure~\ref{fig:overview}. This section provides a detailed explanation of their definitions and functional roles.

\begin{itemize}[left=0pt]
\item \textit{Road/Weather}: refers to the road type or weather conditions in the original scenario dataset and requires a corresponding textual description for prompt generation. Additionally, the \textit{Realism and Robustness Analysis} module conducts evaluation and analysis based on different road types and weather conditions.
\item \textit{Time Step}: defines the number of time steps included in a scenario, which facilitates prompt generation and LLM-based evaluation.
\item \textit{Parameter}: specifically refers to the configuration of NPC vehicles, such as position, rotation, and velocity, which is primarily used for applying numerical mutations and analyzing the impact of individual parameters.
\item \textit{Scenario Number}: indicates the number of driving scenarios that will be used for evaluation.
\item \textit{Mutation Seed}: defines the degree of numerical mutation and can take either positive or negative values.
\item \textit{LLM Configuration}: includes hyperparameter settings related to the LLM, such as the specific LLM API, temperature, and other relevant options.
\item \textit{\# of Repetitions}: indicates the number of repetitions the LLM evaluates a given scenario to ensure the accuracy of the results.
\item \textit{File Configuration}: defines the dataset path, report storage location, and related output options to support file management.
\end{itemize}

\begin{figure}[!t]
    \centering
    \fbox{\includegraphics[width=0.8\linewidth]{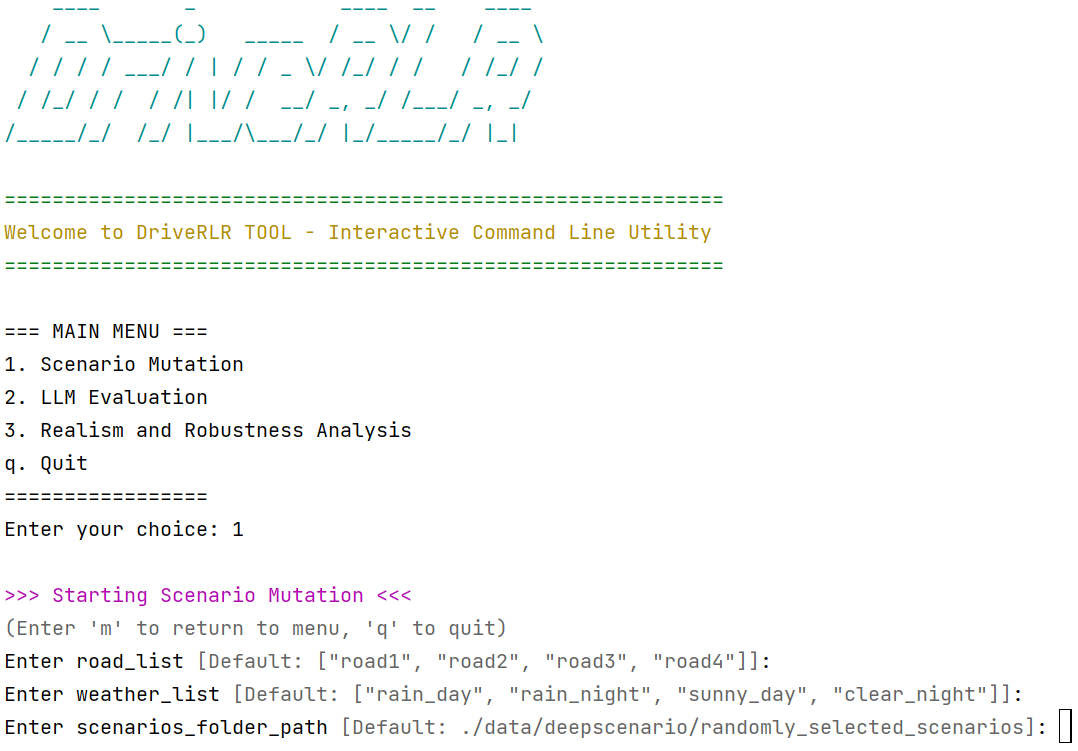}}
    \caption{Interactive Command-line Tool of \method.}
    \label{fig:tool}
\end{figure}

\subsection{Usage Mode}
\noindent
\method provides three usage modes, designed to accommodate users with different levels of functional requirements and technical expertise:

\begin{itemize}[left=0pt]
\item Mode 1: \method offers a fully packaged interactive command-line tool, as shown in Figure~\ref{fig:tool}. Users can select the module they wish to execute and fill in the required configuration parameters to run the tool. The tool also provides default parameter templates within the input interface for reference or modification. This mode is suitable for general users or beginners who want to quickly get started without modifying the underlying code.
\item Mode 2: \method provides an installable Python package that can be easily downloaded and imported into user projects. This mode enables more fine-grained control over specific functions, such as direct invocation of modules like the \textit{JSON Serializer} shown in Figure~\ref{fig:overview}. It is ideal for developers or researchers who require flexible control over the execution process while leveraging built-in functionalities.
\item Mode 3: \method is fully open-source and includes documentation on how to repackage the source code into a custom Python package. This allows advanced users or researchers with specific needs to modify the codebase, add their own methods, or enhance existing functions. For example, users may customize the scenario mutation logic, modify prompt templates, or introduce new evaluation metrics. This mode supports highly customized experimentation and system-level extension.
\end{itemize}

\subsection{Generated Outputs}
\noindent
\method generates three types of outputs based on the three modules shown in Figure~\ref{fig:overview}: \textit{Scenario Mutation}, \textit{LLM Evaluation}, and \textit{Realism and Robustness Analysis}. These outputs include: a collection of original and mutated scenario variants, multiple evaluation outputs from the LLM, and final empirical analysis results.
Scenario variants are stored in JSON format, LLM evaluation outputs are saved as text files, and empirical analysis results include CSV tables and PNG charts.
The output storage location can be customized by the user or set to a default path recommended by \method. Once the execution is complete, users can access and review all outputs in the specified directory.

\section{Validation}
\noindent
\textbf{Experiment Setting}. To evaluate the usability and effectiveness of \method, we replicated and extended the experiments based on the methodology described in~\cite{wu2024reality}.
The original driving scenarios used for evaluation are consistent with those in~\cite{wu2024reality}, and were randomly sampled from the DeepScenario dataset~\cite{lu2023deepscenario}. This dataset enforces strict realism constraints to ensure the realism of the scenarios and serves as the benchmark dataset for \method.
In the updated experiments, we selected three of the most recent LLMs: \gpt, \llama, and \mistral, using the same configuration settings as in~\cite{wu2024reality}.
In addition, all evaluation metrics used in our experiments are aligned with those in~\cite{wu2024reality}.

\begin{table}[!t]
\centering
\caption{Realism Success Rate (\textit{RSR}) and Robustness Score (\textit{RS}) achieved by LLMs. \textit{All} denotes overall results; \textit{R1-R4} represent Road1-Road4, and \textit{RD, RN, SD, and CN} denote Rainy Day, Rainy Night, Sunny Day, and Clear Night, respectively.}
\label{tab:results}
\resizebox{0.55\linewidth}{!}{
\begin{tabular}{lrrr}
\toprule
\multirow{2}{*}{\textbf{Metric}} & \multicolumn{3}{r}{\textbf{Large Language Model}}                              \\ \cmidrule(r){2-4}
                        & \textbf{\gpt} & \textbf{\llama} & \textbf{\mistral} \\ \midrule
\textit{RSR}                     & 61.27\%             & 79.73\%               & 73.64\%                 \\ \midrule
$RS_{All}$                 & 10.74               & 12.37                 & 12.90                   \\ \midrule
$RS_{R1}$                  & 16.87               & 11.75                 & 17.05                   \\
$RS_{R2}$                  & 11.17               & 12.39                 & 12.25                   \\
$RS_{R3}$                  & 7.16                & 11.76                 & 13.19                   \\
$RS_{R4}$                  & 7.77                & 13.59                 & 9.13                    \\ \midrule
$RS_{RD}$                  & 11.28               & 12.70                 & 12.88                   \\
$RS_{RN}$                  & 8.66                & 12.75                 & 13.14                   \\
$RS_{SD}$                  & 13.39               & 12.30                 & 13.73                   \\
$RS_{CN}$                  & 9.64                & 11.73                 & 11.86                  \\ \bottomrule
\end{tabular}
}
\end{table}

\textbf{Results}. Table~\ref{tab:results} presents key evaluation results, including the Realism Success Rate (\textit{RSR}) and Robustness Score (\textit{RS}). Due to space limitations, not all results are shown here. However, the complete evaluation results, along with corresponding intermediate outputs, are available in our public repository~\cite{DriveRLR}.
As shown in Table~\ref{tab:results}, all three LLMs achieved the \textit{RSR} above 60\%. \textit{RSR} metric reflects whether an LLM can accurately determine the realism of driving scenarios. The results indicate that LLMs possess a certain level of capability in assessing scenario realism.
\textit{RS} ranges from -5 to 20, with higher values indicating stronger robustness in evaluating the driving scenario realism. The $RS_{All}$ results in Table~\ref{tab:results} represent the overall robustness of each LLM across all scenarios. Among three LLMs, \mistral achieved the highest robustness (12.90), followed by \llama (12.37), while \gpt performed the weakest (10.74). It means that \mistral is generally more consistent and confident in realism evaluations.
$RS_{R1}$ -- $RS_{R4}$ results break down LLM robustness by different road types. The results show that road type affects LLMs' robustness in assessing scenario realism. For example, \gpt achieved the \textit{RS} of 16.87 on \textit{Road 1}, but only 7.16 on \textit{Road 3}. Similar variations are observed with \llama and \mistral. Moreover, for the same road type, different LLMs do not follow a consistent trend, i.e., none of the LLMs consistently outperformed or underperformed across all roads. This further indicates that road type has a notable impact on LLM robustness in realism evaluation.
$RS_{RD}$, $RS_{RN}$, $RS_{SD}$, and $RS_{CN}$ examine robustness under different weather conditions. Similar to the results observed for different road types, weather conditions also influence the robustness of each LLM. Notably, \mistral consistently achieved the highest \textit{RSs} across all weather conditions, while \gpt and \llama exhibited more variability. This means that \mistral may be less sensitive to weather condition changes and more confident in evaluating scenario realism under varying weather conditions.

\textbf{Discussion}. The experimental conclusions are largely consistent with those reported in~\cite{wu2024reality}, indicating that \method is reliably implemented. Furthermore, by re-running the experiments with the latest LLMs, we demonstrate that \method can easily adapt to different models, highlighting its practical value and broad applicability.
Moreover, \method can support other tasks in ADS testing. For example, its \textit{RS} can be used as an objective function for scenario generation, to promote more realistic scenario creation. It can also help assess simulator realism by comparing \textit{RS} on the same scenarios executed in different simulators, or curating realistic scenario subsets from existing datasets. These capabilities make \method a valuable asset in the ADS testing pipeline.

\section{Conclusion and Future Work}
\noindent
We proposed \method, a benchmarking tool for Large Language Models (LLMs) to evaluate their robustness in scenario realism assessment.
\method mutates original scenarios to generate variants, which are then used to benchmark LLMs to repeatedly evaluate their realism via prompt-based assessments, producing an output set for analyzing the LLM's robustness.
We presented the technical details of \method and validated its effectiveness and usability through experiments on the DeepScenario dataset using three state-of-the-art LLMs.
All data, results, and code are publicly accessible via our open repository~\cite{DriveRLR}.
Currently, \method has only been evaluated on the parameter-based DeepScenario dataset. In future work, we plan to support more diverse scenario formats (e.g., image or video).
We also aim to enhance the evaluation by introducing more fine-grained metrics and statistical tests and supporting a broader range of LLMs to improve adaptability.

\section*{Acknowledgments}
This work is supported by the Co-tester project (No. 314544) and the Co-evolver project (No. 286898/F20), funded by the Research Council of Norway.
Aitor Arrieta is part of the Software and Systems Engineering research group of Mondragon Unibertsitatea (IT1519-22), supported by the Department of Education, Universities and Research of the Basque Country.

\bibliographystyle{unsrt}  
\bibliography{references}

@article{zhao2024autonomous,
  title={Autonomous driving system: A comprehensive survey},
  author={Zhao, Jingyuan and Zhao, Wenyi and Deng, Bo and Wang, Zhenghong and Zhang, Feng and Zheng, Wenxiang and Cao, Wanke and Nan, Jinrui and Lian, Yubo and Burke, Andrew F},
  journal={Expert Systems with Applications},
  volume={242},
  pages={122836},
  year={2024},
  publisher={Elsevier}
}

@article{tang2023survey,
  title={A survey on automated driving system testing: Landscapes and trends},
  author={Tang, Shuncheng and Zhang, Zhenya and Zhang, Yi and Zhou, Jixiang and Guo, Yan and Liu, Shuang and Guo, Shengjian and Li, Yan-Fu and Ma, Lei and Xue, Yinxing and others},
  journal={ACM Transactions on Software Engineering and Methodology},
  volume={32},
  number={5},
  pages={1--62},
  year={2023},
  publisher={ACM New York, NY}
}

@article{zhang2022finding,
  title={Finding critical scenarios for automated driving systems: A systematic mapping study},
  author={Zhang, Xinhai and Tao, Jianbo and Tan, Kaige and T{\"o}rngren, Martin and S{\'a}nchez, Jos{\'e} Manuel Gaspar and Ramli, Muhammad Rusyadi and Tao, Xin and Gyllenhammar, Magnus and Wotawa, Franz and Mohan, Naveen and others},
  journal={IEEE Transactions on Software Engineering},
  volume={49},
  number={3},
  pages={991--1026},
  year={2022},
  publisher={IEEE}
}

@inproceedings{nalic2020scenario,
  title={Scenario based testing of automated driving systems: A literature survey},
  author={Nalic, Demin and Mihalj, Tomislav and B{\"a}umler, Maximilian and Lehmann, Matthias and Eichberger, Arno and Bernsteiner, Stefan},
  booktitle={FISITA web Congress},
  volume={10},
  year={2020}
}

@inproceedings{calo2020generating,
  title={Generating avoidable collision scenarios for testing autonomous driving systems},
  author={Cal{\`o}, Alessandro and Arcaini, Paolo and Ali, Shaukat and Hauer, Florian and Ishikawa, Fuyuki},
  booktitle={2020 IEEE 13th International Conference on Software Testing, Validation and Verification (ICST)},
  pages={375--386},
  year={2020},
  organization={IEEE}
}

@article{lu2024epitester,
  title={Epitester: Testing autonomous vehicles with epigenetic algorithm and attention mechanism},
  author={Lu, Chengjie and Ali, Shaukat and Yue, Tao},
  journal={IEEE Transactions on Software Engineering},
  year={2024},
  publisher={IEEE}
}

@article{feng2023dense,
  title={Dense reinforcement learning for safety validation of autonomous vehicles},
  author={Feng, Shuo and Sun, Haowei and Yan, Xintao and Zhu, Haojie and Zou, Zhengxia and Shen, Shengyin and Liu, Henry X},
  journal={Nature},
  volume={615},
  number={7953},
  pages={620--627},
  year={2023},
  publisher={Nature Publishing Group UK London}
}

@article{liao2025advancing,
  title={Advancing autonomous driving system testing: Demands, challenges, and future directions},
  author={Liao, Yihan and Zhang, Jingyu and Keung, Jacky and Xiao, Yan and Dai, Yurou},
  journal={Information and Software Technology},
  pages={107859},
  year={2025},
  publisher={Elsevier}
}

@article{stocco2022mind,
  title={Mind the gap! a study on the transferability of virtual versus physical-world testing of autonomous driving systems},
  author={Stocco, Andrea and Pulfer, Brian and Tonella, Paolo},
  journal={IEEE Transactions on Software Engineering},
  volume={49},
  number={4},
  pages={1928--1940},
  year={2022},
  publisher={IEEE}
}

@article{lu2023deepqtest,
  title={Deepqtest: testing autonomous driving systems with reinforcement learning and real-world weather data},
  author={Lu, Chengjie and Yue, Tao and Zhang, Man and Ali, Shaukat},
  journal={arXiv preprint arXiv:2310.05170},
  year={2023}
}

@article{yan2023learning,
  title={Learning naturalistic driving environment with statistical realism},
  author={Yan, Xintao and Zou, Zhengxia and Feng, Shuo and Zhu, Haojie and Sun, Haowei and Liu, Henry X},
  journal={Nature communications},
  volume={14},
  number={1},
  pages={2037},
  year={2023},
  publisher={Nature Publishing Group UK London}
}

@article{achiam2023gpt,
  title={Gpt-4 technical report},
  author={Achiam, Josh and Adler, Steven and Agarwal, Sandhini and Ahmad, Lama and Akkaya, Ilge and Aleman, Florencia Leoni and Almeida, Diogo and Altenschmidt, Janko and Altman, Sam and Anadkat, Shyamal and others},
  journal={arXiv preprint arXiv:2303.08774},
  year={2023}
}

@inproceedings{wu2024reality,
  title={Reality bites: Assessing the realism of driving scenarios with large language models},
  author={Wu, Jiahui and Lu, Chengjie and Arrieta, Aitor and Yue, Tao and Ali, Shaukat},
  booktitle={Proceedings of the 2024 IEEE/ACM First International Conference on AI Foundation Models and Software Engineering},
  pages={40--51},
  year={2024}
}

@inproceedings{lu2023deepscenario,
  title={Deepscenario: An open driving scenario dataset for autonomous driving system testing},
  author={Lu, Chengjie and Yue, Tao and Ali, Shaukat},
  booktitle={2023 IEEE/ACM 20th International Conference on Mining Software Repositories (MSR)},
  pages={52--56},
  year={2023},
  organization={IEEE}
}

@misc{gpt-5,
title = {{GPT-5}},
author = {OpenAI},
year = {2025},
howpublished = {\url{https://platform.openai.com/docs/models/gpt-5}},
note = {[Online; accessed 2025-08-25]}
}

@misc{DriveRLR,
title = {{DriveRLR}},
author = {Jiahui Wu},
year = {2025},
howpublished = {\url{https://github.com/Simula-COMPLEX/DriveRLR}},
note = {[Online; accessed 2025-11-05]}
}

\end{document}